\documentclass[preprint,aps,altaffilletter,showkeys]{revtex4} 
\usepackage{graphicx}
\usepackage{color}
\usepackage[colorlinks=true]{hyperref}
\usepackage{bm}
\usepackage{xspace}
\usepackage{times}
\usepackage{graphics}

\newcommand{\ie}{{\it i.e.\,}}

\newcommand{\vs}{{\it vs.\,}}

\begin{document}

\title{Two tone response of radiofrequency signals using the voltage output of a Superconducting Quantum Interference Filter}

\author{P.~Caputo}
\author{J.~Tomes}
\author{J.~Oppenl{\"a}nder}
\author{Ch.~H{\"a}ussler}
\author{A.~Friesch}
\author{T.~Tr\"auble}

\draft

\affiliation{QEST, Quantenelektronische Systeme GmbH, Steig{\"a}ckerstr. 13, 72768 Reutlingen (Germany) }

\author{N.~Schopohl}
\email{nils.schopohl@uni-tuebingen.de}
\thanks{Tel. +49 7071 2978635}
\affiliation{Lehrstuhl f\"ur Theoretische Festk{\"o}rperphysik, Universit{\"a}t T{\"u}%
bingen\\
Auf der Morgenstelle 14, 72076 T{\"u}bingen (Germany) }
\date{\today }

\begin{abstract}

In the presence of weak time harmonic electromagnetic fields, 
Superconducting Quantum Interference Filters (SQIFs) show the typical behavior of non linear mixers. The SQIFs are manufactured from high-$T_{c}$ grain boundary Josephson junctions and
operated in active microcooler. The dependence of dc voltage output $V_\mathrm{dc}$
vs. static external magnetic field $B$ is non-periodic and consists of a
well pronounced unique dip at zero field, with marginal side modulations at
higher fields. We have successfully exploited the parabolic shape of the
voltage dip around $B=0$ to mix \emph{quadratically} two external time
harmonic rf-signals, at frequencies $f_1$ and $f_2$ below the Josephson frequency $f_\mathrm{J}$, and detect the corresponding mixing signal at $%
\mid {f_{1}-f_{2}}\mid $. When the mixing takes place on the SQIF current-voltage characteristics    
the component at $2f_2-f_1$ is present. 
The experiments suggest potential applications of a SQIF as a
non-linear mixing device, capable to operate at frequencies from dc to few
GHz with a large dynamic range.

\end{abstract}


\maketitle
\section*{1. INTRODUCTION}

Superconducting Quantum Interference Filters (SQIFs) are serial (or parallel) arrays of Josephson junctions with an irregular grating structure, \ie a specially selected distribution of the loop areas, such that the phase sensitive superposition of the individual loop modulations leads to a dc voltage response $V_\mathrm{dc}$ \vs static magnetic field $B$ [$V_\mathrm{dc}(B)$- curve] that has a unique dip around $B=0$, with a vanishing modulation everywhere else \cite{Oppenlander:PRB,Haussler:JAP2001}. The detailed shape of the voltage output $V_\mathrm{dc}(B)$, and in particular the dip width $\Delta B$ and its voltage span $\Delta V$, can be engineered by choosing both suitable areas of the loop sequence and 
electric parameters of the Josephson junctions. However, the SQIF performance is not
degraded by the large spread of parameters typically present in high $T_\mathrm{c}$-junctions.
The sensitivity of a SQIF in response to an applied magnetic field is
determined by the \textsl{voltage-to-magnetic field} transfer factor $%
V_\mathrm{B}=\max (\partial V_\mathrm{dc}/\partial B)$. Employing various flux focusing
structures together with a SQIF it is possible to significantly enhance the
transfer factor \cite{Schultze:SST_2003,Schultze:SST_2005}. A flux focusing structure
integrated together with a SQIF array is in the following referred to as a
SQIF-sensor [for a schematic view see Fig.\thinspace \ref{chip}(c)] .

In the resistive mode, the SQIF dc voltage output $V_\mathrm{dc}(I_\mathrm{b}, B)$ is set by the bias
current $I_\mathrm{b}$ and by the static field $B$, and it is the time average of a fast signal  
$V(I_\mathrm{b},B,t) \propto \partial \varphi/\partial t$, where $\varphi$ is the
Josephson phase across the JJ. 
The main harmonic of $V(I_\mathrm{b},B,t)$ has frequency $f_J = V_\mathrm{dc}(I_\mathrm{b}, B)/\Phi_0$ (the second Josephson relation)\cite{BP}. 

Let us consider the case when in addition to the static magnetic field $B$ applied perpendicular to the SQIF loops, there is a weak radiofrequency magnetic field $b_\mathrm{rf}\left( t\right)$, such that $B(t) = B + b_\mathrm{rf}(t)$.  The time harmonic magnetic field is accompanied by a time harmonic electric field, $E(t) = E + e_\mathrm{rf}(t)$. If we assume the electric field to be oriented parallel to the bias leads, the corresponding fluctuations induced in the bias current can be written as $I(t)=I_\mathrm{b} + i_\mathrm{rf}(t)$, which are naturally dispersive due to the frequency dependence of the surface impedance in a superconductor. Thus, if the fields vary slowly in comparison with $f_{J}$, the SQIF voltage can be written as 
\begin{eqnarray}
\label{Eq:Vdc(t)}
&& \left\langle V(\overbrace{I_\mathrm{b} + i_\mathrm{rf}(t)}^{\mathrm {bias \,\, coupling}}, 
\underbrace{B + b_\mathrm{rf}(t)}_{\mathrm {flux \,\, coupling}},t) \right\rangle  = 
\\
&=&  V_\mathrm{dc}(I_\mathrm{b}, B)\,+\, (i_\mathrm{rf}(t) \, \partial_\mathrm{I_b} \,+\, b_\mathrm{rf}(t)\, \partial_\mathrm B) V_\mathrm{dc}(I_\mathrm{b}, B)  
  \,+\,  \frac12 (i_\mathrm{rf}(t) \, \partial_\mathrm{I_{b}} \,+\, b_\mathrm{rf}(t) \, \partial_B)^2 V_\mathrm{dc}(I_\mathrm{b}, B)  + \ldots \nonumber  
\end{eqnarray}
where $\partial_\mathrm{I_{b}}$[$\partial_\mathrm B$] denotes the derivative with respect to $I_\mathrm{b}$ [$B$], and $\langle \rangle$  the time average over the time scale set by $f_J^{-1}$, while the slow time dependence remains. In Eq.~(\ref{Eq:Vdc(t)}) we have denoted as {\em bias coupling} the dependence of the SQIF voltage on the fluctuating induced rf-current, and as {\em flux coupling} the dependence on the fluctuating magnetic flux. For $b_\mathrm{rf}(t) = b_\mathrm{rf}\cos(2\pi f t)$, with amplitude $b_\mathrm{rf} \ll \Delta B$ and  
frequency $f<\frac{f_J}{2}$, Eq.~(\ref{Eq:Vdc(t)}) describes a superposition
of the dc voltage output $V_\mathrm{dc}(I_\mathrm{b}, B)$ 
with slowly varying \textit{rf} terms in $f$ (voltage outputs at frequency $f$, due to bias and flux variations, with pre-factors proportional respectively to the amplitudes $i_\mathrm{rf}$ and $b_\mathrm{rf}$); plus \textit{rf} terms in $2f$, etc...  Thus, the amplitude of the first harmonic of the Fourier spectrum of the output signal contains information about the slopes $\partial_\mathrm B V_\mathrm{dc}\vert_\mathrm{I_b}$
and $\partial_\mathrm{I_b} V_\mathrm{dc}\vert_\mathrm{B}$, while the amplitude of the second harmonic is proportional to the curvature in the $\partial^2_\mathrm B V_\mathrm{dc}\vert_\mathrm{I_b}$ and in the $\partial^2_\mathrm{I_b} V_\mathrm{dc}\vert_\mathrm{B}$ .

The Fourier transform of $V(I_\mathrm{b} + i_\mathrm{rf}(t), B+b_\mathrm{rf}(t),t)$ is the \emph{spectral voltage output} $\widehat{V}(I_\mathrm{b}, B,f)$. Using a spectrum analyzer, the power spectrum $\left\vert \widehat{V}\left( I_\mathrm{b}, B,f\right) \right\vert ^{2}$
can be detected at frequencies around the center
frequency $f$, as a function of the sweeping parameter which can be either $I_\mathrm{b}$ (at a constant value of the static field
$B$), or as a function of $B$ (at constant value of the bias current $I_\mathrm{b}$). If the amplitude of the incident \textit{rf} signal is sufficiently small, 
$b_\mathrm{rf}$ $\ll $ $\Delta B$, the spectral voltage output will be proportional to $\partial_\mathrm B V_\mathrm{dc}\vert_\mathrm{I_b}$ and 
to $\partial_\mathrm{I_b} V_\mathrm{dc}\vert_\mathrm{B}$: it will be
maximum in field regions of maximum slope, 
while in regions of a flat $V_\mathrm{dc}(I_\mathrm{b}, B)$ curves, it will tend to zero.

Quadratic mixing is expected for two tone experiments, where the incident 
\textit{rf} signal is a superposition of two time harmonic signals: $%
b_{\rm {rf}}\left( t\right) =b_\mathrm{rf1}\cos \left( 2\pi f_{1}\,t\right) +b_\mathrm{rf2}\cos
\left( 2\pi f_{2}\,t\right)$ (we assume $f_{1}<f_{2}<\frac{f_{J}}{2}$).
Taking into account the parabolic shape of the $V_\mathrm{dc}(B)$- curve around $%
B=0$, one expects in that region a larger amplitude for the quadratic
mixing signals at frequencies $f_{2}-f_{1}$, $2f_{1}$, $f_{2}+f_{1}$ and $%
2f_{2}$, respectively, compared to the field regions of zero curvature of  
$V_\mathrm{dc}(B)$-characteristics. In a similar manner, in the non linear regions of the $V_\mathrm{dc}(I_\mathrm{b})$- curve, quadratic mixing can also take place. However, in the case of the $V_\mathrm{dc}(I_\mathrm{b})$-curve the spectral voltage output contains higher order mixing products, due to higher order non linearities present in the characteristics.  

Recently, we have demonstrated quadratic mixing of radio frequency signals using SQIF-sensors \cite{Caputo:APL_2006}: the external \textit{rf} signals couple through the time dependent magnetic flux threading the loops of the interferometer, giving rise to a beat-pattern of the original high frequency voltage output (at the Josephson frequency), set by $I_\mathrm{b}$ and $B$. The quadratic mixing takes place due to exact parabolic shape of the $V_\mathrm{dc}(B)$- curve, and the mixed signal is detected at frequencies $\mid {f_{1} \pm f_{2}}\mid $. Due to limited bandwidth of the set-up, the incident signal frequencies were always chosen such that the quadratically mixed components were at a few hundred MHz.
The experiments were performed in active micro-coolers, without significant degradation of the SQIF-sensors, even when the  magnetic shield was removed.

In this paper, we present experiments in which the mixing of radiofrequency signals on the voltage output of the SQIF is extended to larger frequencies.  
The mixing is studied in two ways: either using as sweeping parameter the static magnetic field $B$ (\ie mixing made on the $V_\mathrm{dc}(B)$-  curve, at a constant value of the bias current $I_\mathrm{b}$), or the bias current $I_\mathrm{b}$ (\ie mixing made on the $V_\mathrm{dc}(I_\mathrm{b})$- curve, at $B=0$). As expected, also when the incident fields have larger frequencies the detected spectral voltage output of
our SQIF sensor shows quadratic components, with maximum amplitude in the regions of maximum curvature of the $V_\mathrm{dc}(B)$ curve. 
In the regions of non linear $V_\mathrm{dc}(I_\mathrm{b})$- curve, the quadratic mixing effect takes place as well. This can be explained by the dependence of the SQIF output voltage on the time varying electric field [through the bias current, as described by Eq.(1)]. Only when mixing on the $V_\mathrm{dc}(I_\mathrm{b})$- curve, high order mixing components are also observed. These are never present when mixing on the $V_\mathrm{dc}(B)$- curve - as long as small amplitude signals are used ($b_\mathrm{rf} \ll \Delta B$).

\section*{2. THE SETUP}

\begin{figure}[!tb]
\centering
\includegraphics*{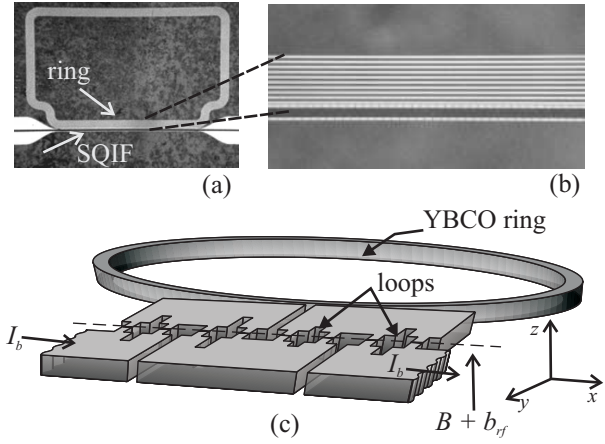} 
\caption{(a) Optical microscope image of the chip with
the SQIF placed along the grain boundary, and the superconducting ring inductively
coupled to the SQIF; (b) zoom on the ring made of 10 equidistant parallel loops, one aligned inside the other. All superconducting
parts, except the regions across the junctions, are covered by gold.
(c) Sketch, not in scale, of the SQIF-sensor, indicating the grain boundary (dashed line), the direction of the bias current and of the magnetic field.}
\label{chip}
\end{figure}

Experiments have been performed with serial SQIFs of 211 loops, consisting
of YBa$_2$Cu$_3$O$_{7-x}$\xspace grain boundary Josephson junctions grown on
24$^{\circ}$- oriented bicrystal MgO substrates\cite{IPHT_Jena}. 
The Josephson junctions are designed with a width of $2\,%
\mathrm{\mu m} $, the YBa$_2$Cu$_3 $O$_{7-x}$\xspace layer being $130\,%
\mathrm{nm}$ thick, so that the resulting junction critical current density
is $J_\mathrm{c} \approx 23 \,\mathrm{kA/cm^2} $, at $T=77\,\mathrm{K}$. The serial loops have an area distribution ranging 
between $38\mathrm{\mu m^{2}}$ and $210\,\mathrm{\mu m^{2}%
}$. In order to enhance the coupling of the SQIF loops to the magnetic flux, we
fabricated a SQIF-sensor with a flux focusing structure, 
\ie\thinspace a 
superconducting ring placed on one side of the grain boundary. Figure\thinspace \ref{chip}(a) is an optical microscope image of the sample. The focusing loop is effectively a split loop design\cite%
{Ludwig:IEEE2001}, consisting of 10 equidistant parallel thin loops, one
aligned inside the other; a fragment of the split loop along the SQIF is enlarged in Fig.\thinspace \ref{chip}(b).
Figure\thinspace \ref{chip}(c) shows a sketch of the SQIF-sensor with the serial loops of the interferometer along the grain boundary, and the inductively coupled flux focusing YBa$_2$Cu$_3 $O$_{7-x}$\xspace ring.  
When the pick-up ring is split in various parallel loops, we have obtained an enhanced sensitivity of the SQIF to the magnetic field (larger transfer factor) with respect to the case of a single broad ring. Calculations show \cite{Ludwig:IEEE2001} that the split ring design leads, in average, to a larger current density compared to a single ring of the same cross section.

The static magnetic field $B$ is applied via a
multi-turn coil placed inside the cooler. The \textit{rf} magnetic field $%
b_\mathrm{rf}(t)$ is applied via a $50\,\mathrm{\Omega }$-loop antenna (primary antenna). Experiments
are made either with a mu-metal shield - in this case the \textit{rf}
antenna was placed inside the shield at a distance of about $5\,\mathrm{cm}$
from the chip -, or with an unshielded cryocooler in open space.
%
%

Samples have been operated in a Stirling microcooler \cite{AIM}, at
temperatures from 55 to $82 \,\mathrm{K}$. Active cooling does not degrade
the SQIF performance, and offers numerous advantages, such as the possibility to set stable temperatures over a large temperature range, quick thermal cycles, compact and portable setups 
\cite{Caputo:APL_2004,Oppenlander:ASC_2005}. To provide transparency to the
incoming \textit{rf} fields, a glass chamber surrounds the microcooler cold head. The sample is  
anchored to the cold head only from one side, in order to keep the focusing ring as much as possible free of
metallic ground plane. On the cold head, an \textit{rf} amplifier is also mounted,
designed for cryogenic applications with a gain of about $40 \,\mathrm{dB}$,
for the bandwidth $0.04 - 6 \,\mathrm{GHz}$. At room temperature and at the typical bias values, we have measured a noise figure of the amplifier of about 1 dB, in a frequency range from 100 MHz to 1 GHz. The amplifier input resistance
is $5 \,\mathrm{k\Omega}$. 
The SQIF is coupled to the amplifier through a coplanar line, after which a bias-Tee is placed. The bias-Tee splits the ac and the dc parts of
the SQIF signal: the ac part, through a capacitor, follows the input line to
the first stage of the cold amplifier; the dc part (bias input \& voltage
output), through a conical inductor, goes to the room temperature electronics,
via twisted wires. 
The amplifier output is carried outside the cooler via an SMA feed-through and flexible coaxial
lines, and detected by a spectrum analyzer. 

The \textit{rf} measurements are performed in the following way:
the spectrum analyzer is operated in \textsl{Zero Span Mode}, \ie as a narrowband receiver tuned at the central frequency $f_\mathrm{0}$ (equal to the frequency at which the detection of the SQIF spectral voltage output is made), and with
a bandwidth around $f_\mathrm{0}$ set by the Resolution Bandwidth (ResBW). To this mode, it corresponds an analog output proportional to the maximum amplitude of the  signal at $f_\mathrm{0}$. This output voltage is recorded by computer while slowly sweeping either the static
magnetic field $B$ (simultaneously, the $V_\mathrm{dc}(B)$- curve is acquired), or sweeping the static bias current $I_\mathrm{b}$ (simultaneously, the $V_\mathrm{dc}(I_\mathrm{b})$- curve is acquired).

\section*{3. THE EXPERIMENTS}
\subsection*{3.1 Mixing on the SQIF $V_\mathrm{dc}(B)$- curve}

We first measure the $%
V_\mathrm{dc}(B)$ dependence of the SQIF at a constant $I_\mathrm{b}$ value, and check the dip symmetry with respect
to zero magnetic field. There is an optimal temperature at which the voltage swing is
a maximum, at a properly chosen bias current. At $T \approx 73\, \mathrm{K}$ and at $I_\mathrm{b} = 20\, 
\mathrm{\mu A}$, we measure a voltage span $\Delta V \approx 1160\, \mathrm{\mu V}$ and a transfer factor $%
V_\mathrm{B} \approx 6500 \, \mathrm{V/T}$. The SQIF normal resistance is $R = 186\,
\Omega $. Successively, the
\textit{rf} is switched ON: a time harmonic signal which is the superposition of two \textit{rf} signals with frequencies $f_\mathrm{1}$ and $f_\mathrm{2}$ and 
equal amplitudes $b_\mathrm{rf1}$ and $b_\mathrm{rf2}$ is applied to the primary antenna, while the \textit{rf} output of the SQIF is
detected. 
A small amplitude of the incoming signals is required, so that the \textit{rf} signal superimposed to the static field does not
modulated the SQIF working point out of the dip and corrupt the effect; but
it has to be high enough so that the SQIF spectral voltage output is above the noise level
set by the resolution bandwidth (ResBW) of the spectrum analyzer. 
The analog output of the spectrum analyzer is recorded by computer while slowly sweeping the static field $B$
(sweep frequency in the kHz range); simultaneously, the \textit{dc} voltage output $V_\mathrm{dc}$ is
acquired as a function of $B$. In the output spectra, we find components at $f_0=\mid {f_1-f_2}\mid$, whose amplitude is maximal at the dip bottom and vanishes in the region of zero curvature of the $V_\mathrm{dc}(B)$ curve. Indeed, the amplitude of the second harmonics (the term in ${f_1-f_2}$) is expected to be proportional to the curvature in the $\partial^2_\mathrm B V_\mathrm{dc}\vert_\mathrm{I_b}$. We have covered the frequency interval from few hundred MHz to few GHz for the two primary incident fields, keeping the difference frequency within the bandwidth of the rf amplifier, for example at values ${f_1-f_2} = 1.5 \mathrm{GHz}$, but also $2$ and $2.5\,\mathrm{GHz}$. 

\begin{figure*}[!htb]
\centering
\includegraphics*{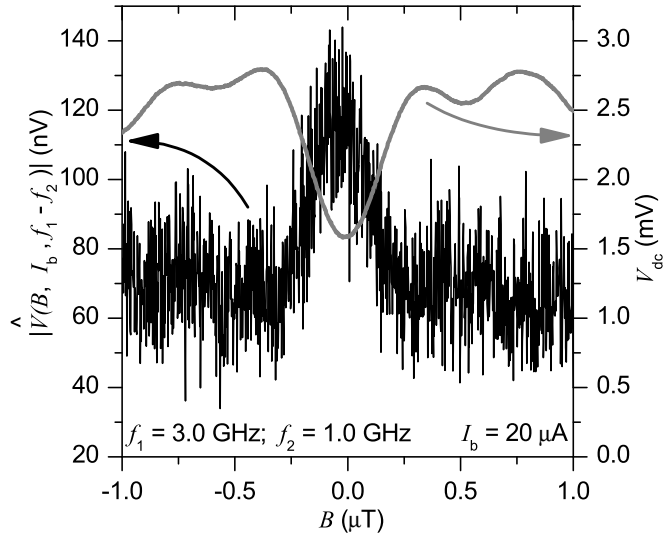}
\caption{The two incident signals are set at $f_{1}=3.0\,\mathrm{GHz}$ and $f_{2}=1.0\,\mathrm{GHz}$. The \textit{rf} voltage is measured with a ResBW of $30\,\mathrm{kHz}$. $I_\mathrm{b}=20\,\mathrm{\mu A}$ and $T=\,73\mathrm{K}$. Gray curve, relative to right axis: $V_\mathrm{dc}$ vs. $B\,$; black curve,
rel. to left axis: \textit{rf} voltage $\left\vert \widehat{V}\left( B, I_\mathrm{b}, f\right) \right\vert$ detected at $f_0=f_{1}-f_{2} = 2\,\mathrm{GHz}$.}
\label{VB}
\end{figure*}

Figures \ref{VB} displays the typically measured curves. 
The incident field is a linear
combination of two signals with frequencies $f_{1}=3\,\mathrm{%
GHz}$ and $f_{2}=1\,\mathrm{GHz}$, and equal amplitudes. In the spectral voltage output of the SQIF, we find then a
quadratic mixing signal at the difference frequency $f_\mathrm{0}=2\,
\mathrm{GHz}$, and in the Figure is displayed the spectral voltage output detected at the central frequency $f_\mathrm{0}$, with ResBW equal to $30\,\mathrm{kHz}$, as a function of the sweeping parameter $B$. The amplitude of the voltage output at $f_\mathrm{0}$ is maximal around $B=0$, and vanishes as the dc working point approaches the dip slopes.

\subsection*{3.1 Mixing on the SQIF $V_\mathrm{dc}(I_\mathrm{b})$- curve}

\begin{figure*}[!htb]
\centering
\includegraphics*{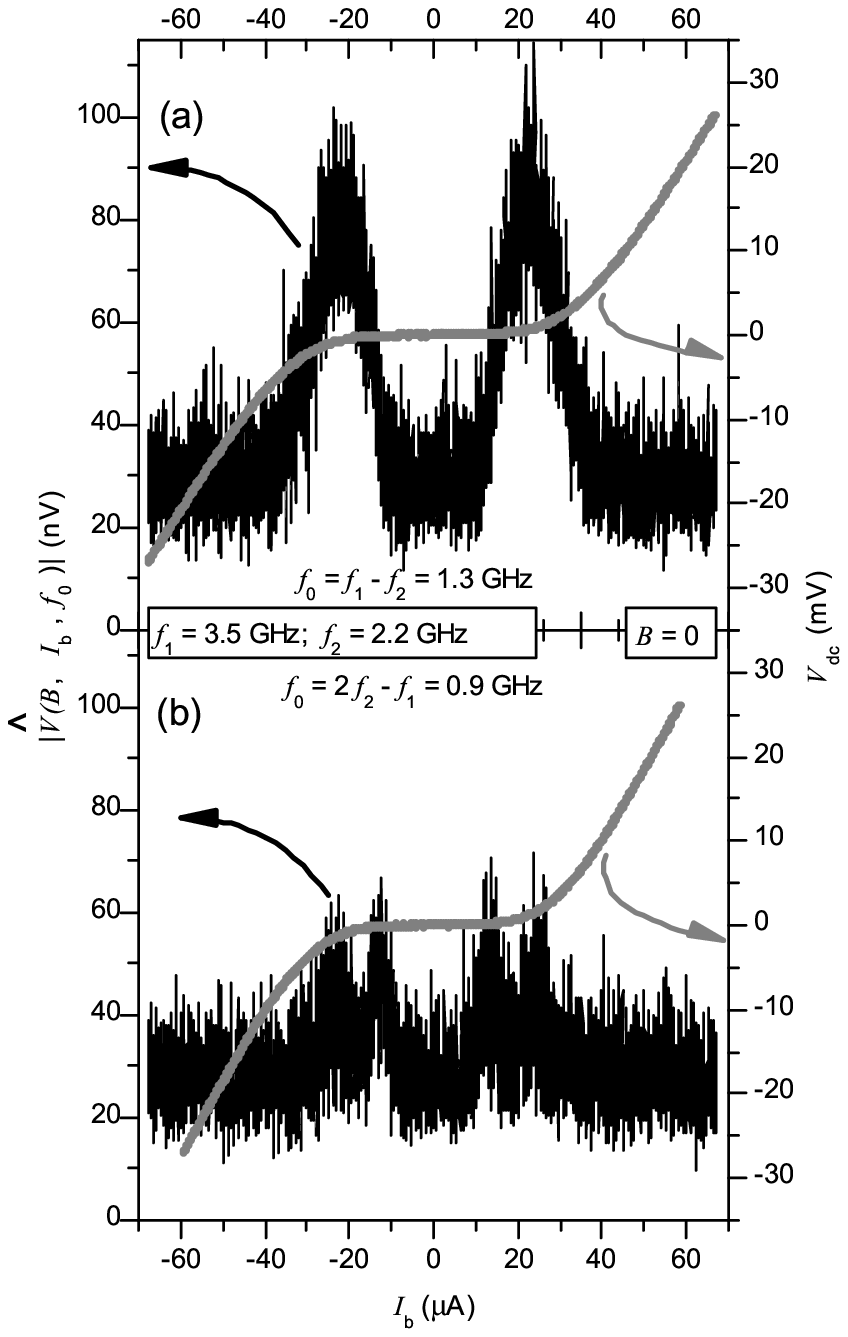}
\caption{The two incident signals are set at $f_{1}=3.5\,\mathrm{GHz}$ and $f_{2}=2.2\,\mathrm{GHz}$. The \textit{rf} voltage is measured with a ResBW of $3\,\mathrm{kHz}$. $B = 0$ and $T=73.5\,\mathrm{K}$. 
(a) Gray curve, relative to right axis: $V_\mathrm{dc}$ vs. $I_\mathrm{b}\,$; black curve,
rel. to left axis: \textit{rf} voltage $\left\vert \widehat{V}\left( B, I_\mathrm{b}, f\right) \right\vert$ detected at $f_\mathrm{0}= {f_{1}-f_{2}} =1.3\,\mathrm{GHz}$; 
(b) Gray curve, relative to right axis: $V_\mathrm{dc}$ vs. $I_\mathrm{b}\,$; black curve,
rel. to left axis: \textit{rf} voltage $\left\vert \widehat{V}\left( B, I_\mathrm{b}, f\right) \right\vert$ detected at $f_\mathrm{0}= 2{f_{2}-f_{1}} = 0.9\,\mathrm{GHz}$.}
\label{IV}
\end{figure*}

At $B=0$, the $V_\mathrm{dc}(I_\mathrm{b})$ is swept; the two incident \textit{rf} signals are applied to the primary antenna, and the spectral voltage output of the SQIF sensor is detected. To the applied time harmonic magnetic field, there corresponds naturally a time harmonic electric field, which couples to the SQIF voltage output through the fluctuations induced in the bias current (if we assume the electric field to have a component parallel to the bias line). Thus, the spectral voltage output detected \vs $I_\mathrm{b}$ might also contain fingerprints of the incident signals, and eventually mix them.  Figures \ref{IV}(a-b) display the typically measured curves. 
The incident field is a linear
combination of two signals with frequencies $f_\mathrm{1}=3.5\,\mathrm{%
GHz}$ and $f_\mathrm{2}=2.2\,\mathrm{GHz}$, and equal amplitudes. In the spectral voltage output of the SQIF, we find the 
quadratic mixing component at the difference frequency $f_\mathrm{0}=f_\mathrm{1}-f_\mathrm{2}=1.3\,
\mathrm{GHz}$. In Fig.\ref{IV}(a) is reported the voltage output detected at $f_\mathrm{0}=1.3\,
\mathrm{GHz}$, with a ResBW of $3\,\mathrm{kHz}$, as a function of the sweeping parameter $I_\mathrm{b}$. Also in this case, the amplitude of the voltage output at $f_\mathrm{1}-f_\mathrm{2}$ is found to be maximal around the region of maximal curvature of the $V_\mathrm{dc}(I_\mathrm{b})$- characteristics. 
In contrast to the case of the $V_\mathrm{dc}(B)$- curve, when sweeping the $V_\mathrm{dc}(I_\mathrm{b})$- curve the spectral voltage output contains also a component at $f_\mathrm{0}= 2f_\mathrm{2}-f_\mathrm{1}$. Figure \ref{IV}(b) shows the SQIF \textit{rf} voltage output detected at $f_\mathrm{0} = 0.9\,\mathrm{GHz}$. 
As expected, while the amplitude of the term in ${f_1-f_2}$ is proportional to the curvature in the $\partial^2_\mathrm{I_{b}} V_\mathrm{dc}\vert_\mathrm{B}$,  
the amplitude of the mixing component at frequency $2f_\mathrm{2}-f_\mathrm{1}$ is proportional to the absolute value of the 3-rd order derivative of the $V_\mathrm{dc}(I_\mathrm{b})$- curve.
 
We have also performed mixing experiments without magnetic shield, both on the $V_\mathrm{dc}(B)$- and on the  $V_\mathrm{dc}(I_\mathrm{b})$- curve. The presence of
environmental disturbances does not suppress the effect (although in this case
the compensation field is different than zero), and the mixing components are clearly visible.

\subsection*{CONCLUSIONS}

In conclusion, we have demonstrated quadratic mixing of radiofrequency signals using the voltage output of a SQIF, both on the $V_\mathrm{dc}(B)$- and on the $V_\mathrm{dc}(I_\mathrm{b})$- curves.  The SQIF sensor is capable to
transfer an incident radiofrequency signal into a converted radiofrequency
voltage signal: the external \textit{rf} fields couple either through the time dependent magnetic flux threading the loops of the interferometer or through the bias current, giving rise to a beat-pattern of the original high frequency voltage output at the Josephson frequency.  In the adiabatic approximation, the quadratic mixing of the 
incident \textit{rf} signals takes place due to the exact parabolic shape of the \textit{dc} voltage output [$V_\mathrm{dc}(B)$- curve] around $B=0$. When mixing on the $V_\mathrm{dc}(I_\mathrm{b})$- curve, in addition to the quadratic components we have found higher order mixing products. In all cases, the detected harmonics have an amplitude dependence proportional to the corresponding order derivative of the $V_\mathrm{dc}$- curve with respect to the sweeping parameter.    

In the presented experiments, by means of a SQIF sensor we have obtained quadratic mixing of incident fields in a frequency range from few hundred MHz up to several GHz, without significant degradation of performance. This feature might open new possibilities for applications of SQIF sensors as broad band mixers.

The authors are very thankful to R.~IJsselsteijn to provide us with samples, and V.~Schultze for many useful discussions.


\begin{references}



\bibitem{Oppenlander:PRB} J.~Oppenl{\"a}nder, Ch.~H{\"a}ussler, and N.~Schopohl, Phys.\ Rev. B, 
{\bf {63}}, 024511 (2000).

\bibitem{Haussler:JAP2001} Ch.~H{\"a}ussler, J.~Oppenl{\"a}nder, and N.~Schopohl, J. Appl. Phys., {\bf{89}}, 1875 (2001).

\bibitem{Schultze:SST_2003} V.~Schultze, R.~IJsselsteijn, R.~Boucher, H.--G.~Meyer, J.~Oppenl{\"a}nder and {Ch. H{\"a}ussler}, and N.~Schopohl, Supercond. Sci. Technol., {\bf{16}}, 1356 (2003). 

\bibitem{Schultze:SST_2005} V.~Schultze, R.~IJsselsteijn and H.--G.~Meyer, Supercond. Sci. Technol., {\bf{19}}, 411 (2006).

\bibitem{Oppenlander:APL_2003} J.~Oppenl{\"a}nder, P.~Caputo, Ch.~H{\"a}ussler, 
T.~Tr\"auble, J.~Tomes, A.~Friesch, and N.~Schopohl, Appl. Phys. Lett., {\bf{83}}, 969 (2003). 

\bibitem{Caputo:APL_2006} P.~Caputo, J.~Tomes, J.~Oppenl{\"a}nder, {Ch. H{\"a}ussler}, A.~Friesch, T.~Tr{\"a}uble, and N.~Schopohl, Appl. Phys. Lett., {\bf{89}}, 062507, (2006).

\bibitem{BP} A.~Barone and G.~Patern{\`o}, {\it Physics and Applications of the Josephson Effect}, J. WILEY \& SONS, New York (1982).

\bibitem{AIM} AIM, AEG Infrarot-Module GmbH, Heilbronn, Germany.

\bibitem{Oppenlander:ASC_2005} J.~Oppenl{\"a}nder, {Ch. H{\"a}ussler}, A.~Friesch, J.~Tomes, P.~Caputo, T.~Tr{\"a}uble, and N.~Schopohl, IEEE Trans. Appl. Supercond., {\bf{15}}, 936 (2005).

\bibitem{Caputo:APL_2004} P.~Caputo, J.~Oppenl{\"a}nder, Ch. H{\"a}ussler, J.~Tomes, A.~Friesch, T.~Tr{\"a}uble, and N.~Schopohl, Appl. Phys. Lett., {\bf{85}}, 1389 (2004).

\bibitem{Caputo:IEEE_2005} P.~Caputo, J.~Tomes, J.~Oppenl{\"a}nder, {Ch. H{\"a}ussler}, A.~Friesch, T.~Tr{\"a}uble, and N.~Schopohl, IEEE Trans. Appl. Supercond., {\bf{15}}, 1044 (2005).
 
\bibitem{IPHT_Jena} Institute for Physical High Technology (IPHT), Jena, Germany.

\bibitem{Ludwig:IEEE2001} F.~Ludwig, A.~B.~M.~Jansman, D.~Drung, M.~O.~Lindstr{\"o}m, S.~Bechstein, J.~Beyer, J.~Flokstra, and T.~Schurig, IEEE Trans. Appl. Supercond., {\bf{11}}, 1315 (2001). 

\end{references}
\end{document}